# Refractive Index-Based Control of Hyperbolic Phonon-Polariton Propagation

*Alireza Fali[†], Samuel T. White[‡], Thomas G. Folland[∥], Mingze He[∥], Neda A. Aghamiri[†], Song Liu[⊥], James H. Edgar[⊥, δ], Joshua D. Caldwell[∥, δ], Richard F. Haglund[‡, δ], Yohannes Abate[\*,†]*

[†] Department of Physics and Astronomy, University of Georgia, Athens, GA 30602, United States

[‡] Department of Physics and Astronomy, Vanderbilt University, Nashville, TN 37235, United States

[∥] Department of Mechanical Engineering, Vanderbilt University, Nashville, TN 37212, United States

[⊥] Tim Taylor Department of Chemical Engineering, Kansas State University, Manhattan, KS 66506 USA

[δ] Interdisciplinary Materials Science Program, Vanderbilt University, Nashville, TN 37212

ABSTRACT: Hyperbolic phonon polaritons (HPhPs) are generated when infrared photons couple to polar optic phonons in anisotropic media, confining long-wavelength light to nanoscale volumes. However, to realize the full potential of HPhPs for infrared optics, it is crucial to understand propagation and loss mechanisms on substrates suitable for applications from waveguiding to infrared sensing. In this paper, we employ scattering-type scanning near-field optical microscopy (s-SNOM) and nano-Fourier transform infrared (FTIR) spectroscopy, in concert with analytical and numerical calculations, to elucidate HPhP characteristics as a function of the complex substrate dielectric function. We consider propagation on suspended, dielectric and metallic substrates to demonstrate that the thickness-normalized wavevector can be reduced by a factor of 25 simply by changing the substrate from dielectric to metallic behavior. Moreover, by incorporating the imaginary contribution to the dielectric function in lossy materials, the wavevector can be dynamically controlled by small local variations in loss or carrier density. Counterintuitively, higher-order HPhP modes are shown to exhibit the same change in polariton wavevector as the fundamental mode, despite the drastic differences in the evanescent ranges of these polaritons. However, because polariton refraction is dictated by the fractional change in the wavevector, this still results in significant differences in polariton refraction and reduced sensitivity to substrate-induced losses for the higher order HPhPs. Such effects may therefore be used to spatially separate hyperbolic modes of different orders, and indicates that for index-based sensing schemes that HPhPs can be more sensitive than surface polaritons in the thin film limit. Our results advance our understanding of fundamental polariton excitations and their potential for on-chip photonics and planar metasurface optics.

Due to the long free-space wavelength of mid- to far-infrared (IR) light, the field of IR nanophotonics has employed polariton effects to confine light to dimensions below the diffraction limit. This has advanced the state of the art in on-chip photonics,[1] polariton waveguides[2] and nanolasers.[3] The hyperbolic polariton offers significant promise in many nanophotonic applications[4] because it offers volume-confined electromagnetic near-fields,[5-8] a restricted propagation angle dictated by the hyperbolic dielectric function,[9,10] and a dramatically expanded photon density of states[6, 10] due to the inclusion of higher-order polaritons with diminishing polariton wavelengths (increasing wavevectors) at the same frequency.[7, 11, 12] Applications of these properties includehyperlensing[5, 13-15], metasurface-based optical components,[16, 17] quantum optics[18] and probes of nanoscale defects.[19, 20]

In 2014 hBN was first reported as a naturally hyperbolic material with exceptionally low optical losses,[12, 21, 22] because it supports polaritons derived from optic phonons[21, 23] rather than scattering from free carriers. Since then, an extensive list of naturally hyperbolic materials have been cataloged;[24, 25] one of the most promising is $MoO_3$[1, 26-28] given its record polariton lifetimes (up to 20 ps)[26] and in-plane hyperbolicity. Unlike surface-confined polaritons,[29] volume-confined hyperbolic polariton fields can interact with the local environment with minimal optical loss. Thus its hyperbolic polariton properties can be tuned without substantially reducing propagation lengths.[2]

Previously we demonstrated planar meta-optics in heterostructures comprising hBN and the phase-change material (PCM) vanadium dioxide ($VO_2$) by showing that refraction of HPhPs in hBN obeys Snell's law at boundaries between rutile and monoclinic domains of $VO_2$.[2] Refraction, induced by a large (1.6x change) wavevector mismatch between principal HPhP modes in hBN supported over the two domains, opens pathways to reconfigurable metasurfaces, rewritable designer planar optics and waveguides, and tunable optical resonators. The volume confinement of the HPhPs guaranteed that although the polaritons are sensitive to the local environment, the polaritonic near-fields will propagate primarily within the low-loss hyperbolic material. This not only provided a first demonstration of a reconfigurable, hyperbolic metasurface, but also illustrated the sensitivity of the hyperbolic polaritons to the local dielectric environment, unlike earlier studies with hBN frustums.[7] The effects of the substrate dielectric function upon HPhPs has to this point received limited investigation. A recent study of HPhPs in suspended hBN by real-space nanoimaging[30, 31] showed that the propagation figure of merit (FoM) was significantly reduced for hBN supported on $SiO_2$.[30] Furthermore, for wrinkled hBN on gold the polariton wavelength is compressed by a factor two.[32, 33] For advanced applications featuring the growing library of natural hyperbolic materials,[24, 25, 34] such as reconfigurable planar optics, the role of the substrate in dictating the room-temperature HPhP dissipation and propagation on the principal and higher-order HPhP modes must be ascertained. Thus, it is critical to quantify how the local dielectric environment of the substrate modifies propagating hyperbolic polariton modes, and to analyze how this propagation is affected by dielectric losses in the substrate.

To this end, we have explored the effects of substrates with varying refractive indices on HPhPs in isotopically enriched hBN, on suspended, metallic, dielectric and phase-change substrates. Employing the analytical model first proposed by Dai et al.,[11] we show that the complex substrate dielectric function has non-trivial consequences for hyperbolic polariton propagation. The model specifically predicts that in the limit of small real substrate permittivity that the hyperbolic polariton wavelength can be modified by up to a factor of 25, simply by transitioning from a metallic to a dielectric substrate. The trends of all hyperbolic polariton wavelengths as a function of the real part of the substrate permittivity are inverted, with metallic (dielectric) substrates exhibiting shrinking (expanding) polariton wavelength with increasing substrate permittivity. Counterintuitively, despite the increasing confinement of the polariton fields to the volume of the hyperbolic medium, the influence of such changes in the substrate permittivity upon the wavelength of the higher-order hyperbolic modes is equivalent to that of the fundamental polariton. Our experiments and calculations also highlight that the imaginary part of the substrate permittivity plays a critical role in dictating hyperbolic polariton propagation. While we report these results using hBN, the findings discussed herein can be generalized to the broader class of hyperbolic media. Based on our findings, we highlight their technological implications:, the various fundamental and higher order modes can be spatially separated and offers promise for index-based sensing modalities by controlling the substrate dielectric function at a local level.

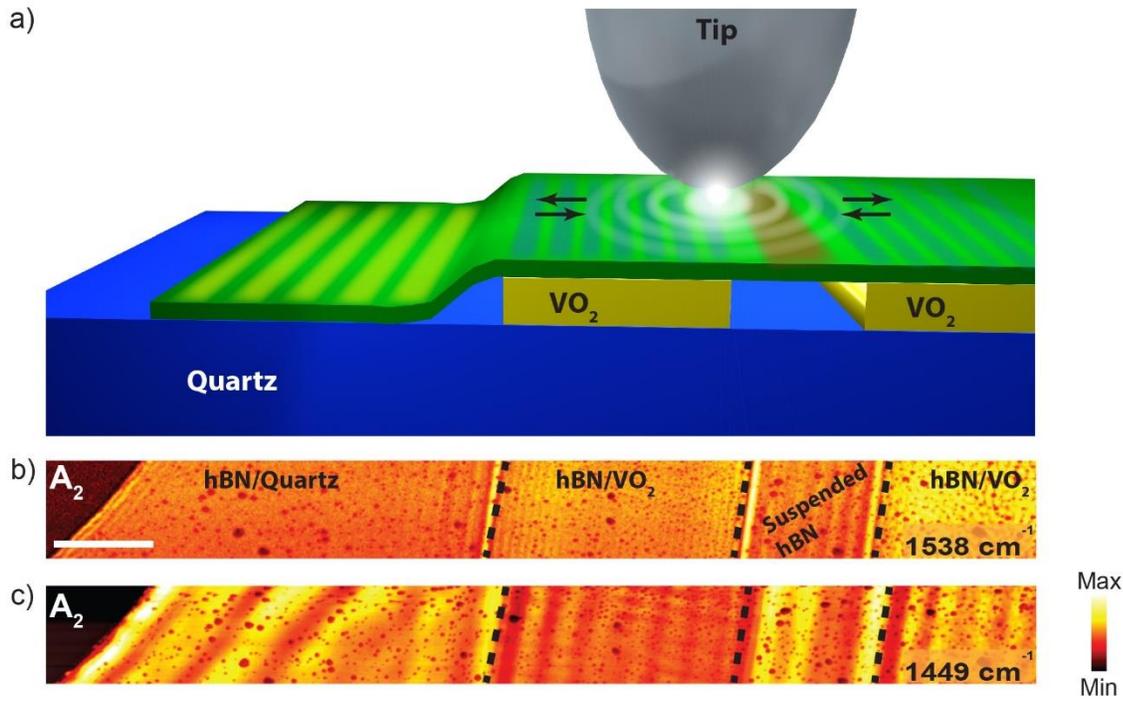

**Figure 1.** Measuring hyperbolic polaritons in different dielectric environments. a) A schematic of the experimental setup, featuring the tip-launched HPhPs on an exfoliated hBN flake (thickness 65 nm) supported by quartz, single-crystal $VO_2$ and suspended in air in the region between two $VO_2$ crystals. The experimental s-SNOM maps of the optical amplitude at b) 1538 cm$^{-1}$ (6.5 µm) and c) 1449 cm$^{-1}$ (6.9 µm) incident frequencies illustrate the propagation of the HPhPs in the hBN flake over these different substrates. The changing periodicity of the optical contrast in each region indicates a modified HPhP wavelength. Scale bar is 5 µm.

To quantify the role of the substrate complex refractive index on the hyperbolic polariton wavevector, we prepared several hBN flakes (see Methods) of similar thicknesses on silicon, quartz, $VO_2$ (insulating and metallic) and silver. We then probed the HPhPs in the hBN using a scattering-type scanning near-field optical microscope (s-SNOM) coupled to a line-tunable quantum cascade laser or a broadband IR source (Methods). When laser light is scattered from the AFM tip, HPhPs are launched by the evanescent fields induced at the tip apex, which is located in near-field proximity to the hBN surface (Figure 1a). These HPhPs propagate radially outward from the tip, within the volume of the hBN flake. Upon reaching a boundary, such as a sharp flake edge or a local domain with significant index contrast, the HPhP is reflected. This reflected polariton wave interferes with the outgoing mode to generate a pattern that can be directly probed by the s-SNOM tip, and for the "tip-launched" polaritons that is half the incident polariton wavelength. Other HPhPs can be directly launched by IR light scattered by the flake edge, which is out-coupled to the detector via the s-SNOM tip. These "edge-launched" modes exhibit a different interference pattern with the periodicity of the incident polariton wavelength.[2,11] Thus, by probing polariton wavelength and propagation length as a function of incident frequency and the complex dielectric constant of the substrate, it is possible to extract quantitatively the substrate-modified HPhP dispersion.

The HPhPs propagating within the hBN flake can be observed in the spatial profiles of s-SNOM amplitude (Figures 1b and c) collected at $\omega$=1538 and $\omega$=1449 cm$^{-1}$. The exfoliated 65 nm thick hBN flake is supported on quartz and draped over two $VO_2$ single crystals, resulting in three regions where the hBN is suspended in air (Figure 1a). The near-field amplitude maps collected at both incident laser frequencies show that the separation between interference fringes, and thus the wavelength of the principal HPhP mode, is strongly modified by the refractive index of the medium over which the mode is propagating, consistent with recent reports.[30] The longest HPhP propagation length is over the suspended region between the two $VO_2$ crystals. The polariton wavelength was similarly substrate dependent in a second hBN flake supported between a Si substrate and $VO_2$ crystal (Figure S1).

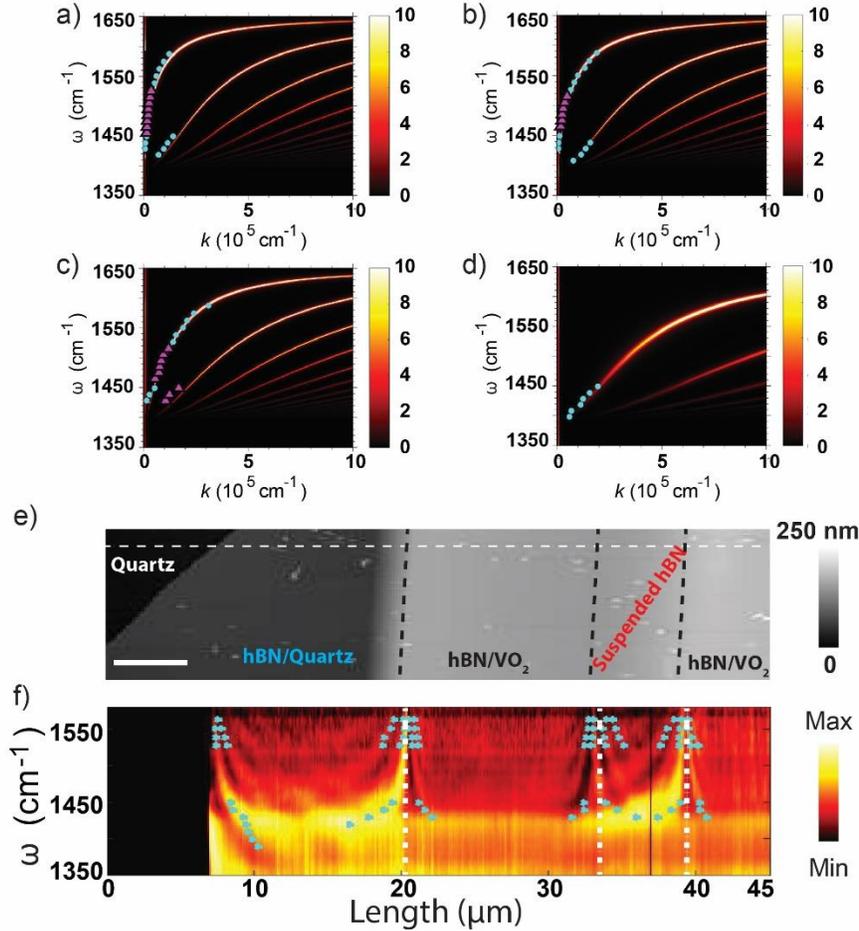

**Figure 2.** Measurements of the dispersion relations of HPhPs in 65 nm thick hBN on different substrates. The polariton wavelengths (ω, wavevectors) for each incident frequency were extracted using fast Fourier transforms of linescans extracted from the s-SNOM amplitude plots for the HPhP modes collected over the various substrates. The dispersion in these values for HPhPs propagating within hBN over a) air (suspended), b) quartz, c) insulating $VO_2$ and d) metallic $VO_2$. Circles indicate data points from s-SNOM and triangles are similar results derived from nano-FTIR spectral linescans (results from f) and are superimposed upon analytical calculations of the HPhP dispersion. The z-scale is the imaginary part of the p-polarized reflection coefficient, Im($r_p$) (e) Topographic maps of the sample illustrate the lack of surface features. (f) Corresponding nano-FTIR spectral linescans illustrate the HPhP properties at every position along the broken white line in e). Solid points in f) are experimental data points extracted from monochromatic polariton images (as in Figures 1b and c).

To capture the dispersion of the propagating HPhPs on different substrates, we collected s-SNOM amplitude images, similar to those in Figures 1b and c, at several laser frequencies. From these images we extracted line-scans parallel to the propagation direction, and then utilized a fast Fourier transform (FFT) to extract the HPhP wavelength ($\lambda_{HPhP}$), then plotted the dependence of this wavelength on laser excitation frequency ($\omega$) for each substrate. We extracted the magnitude of the in-plane wavevector $k$, using $k=2\pi/\lambda_{HPhP}$. The experimental substrate-dependent dispersion relations were then determined by plotting $k$ as a function of $\omega$ (triangular points in Figure 2), for HPhP modes within hBN (a) suspended in air between the two $VO_2$ crystals, (b) on quartz, (c) on insulating $VO_2$ and (d) on metallic $VO_2$.

To supplement these results and provide experimental data in the spectral gap between 1450 to 1480 cm$^{-1}$, we acquired a nano-FTIR linescan that yields the relationship between the momentum $k$ and the excitation frequency $\omega$ at every pixel. The nano-FTIR data were acquired by taking a broadband spectrum along the white dashed line shown in Figure 2e and displaying the resulting spectra in a 2D plot where the x-axis (length) is pixel location and the y-axis is the frequency ($\omega$) covering the polariton spectral range, as shown in Figure 2f. The line-scan covers all of the substrate environments, including suspended hBN, as well as regions where the hBN was in direct contact with quartz and $VO_2$, providing a $k$ vs $\omega$ spatial map over the various substrates in a single scan. This enabled us to extract more data points that would otherwise be missing due to lack of monochromatic laser (in purple triangles in Figures 2a-c).[11] The experimental data are in excellent agreement with the analytical dispersion relations, shown as solid lines for all substrates. These calculations were performed using the analytical model reported in Ref. [11]. In the limit where the HPhP wavelength is much shorter than the wavelength of light in the underlying substrate, this analytic expression for the complex wavevector can be derived from the Fabry-Perot resonance condition:[4]

$$kd = [Re(k) + iIm(k)]d = -\psi \left[ tan^{-1}\left(\frac{\varepsilon_0}{\varepsilon_t \psi}\right) + tan^{-1}\left(\frac{\varepsilon_s}{\varepsilon_t \psi}\right) + \pi l \right], \psi = -i\sqrt{\frac{\varepsilon_z}{\varepsilon_t}} \quad (1)$$

Where $d$ is the hBN thickness, $\varepsilon_0, \varepsilon_s, \varepsilon_t$ and $\varepsilon_z$ are the complex dielectric functions of air, the substrate, and hBN for both in- and out-of-plane directions, respectively, and $l$ is the mode order of the HPhP (0,1,2, …). The three terms in Eq. 1 represent the phase shift accumulated from reflection of the HPhP from the top and bottom of the flake, and during propagation within the layer, respectively.[11] In this approximate equation, the polariton wavelength is normalized to the hBN thickness, facilitating direct comparison of the HPhP dispersion in hBN flakes of varying thickness on different substrates. In addition to the substrates already discussed, similar measurements were made on silverand silicon. The HPhP images and dispersion plots of hBN on these substrates are provided in the Supporting Information, (Figure S2).

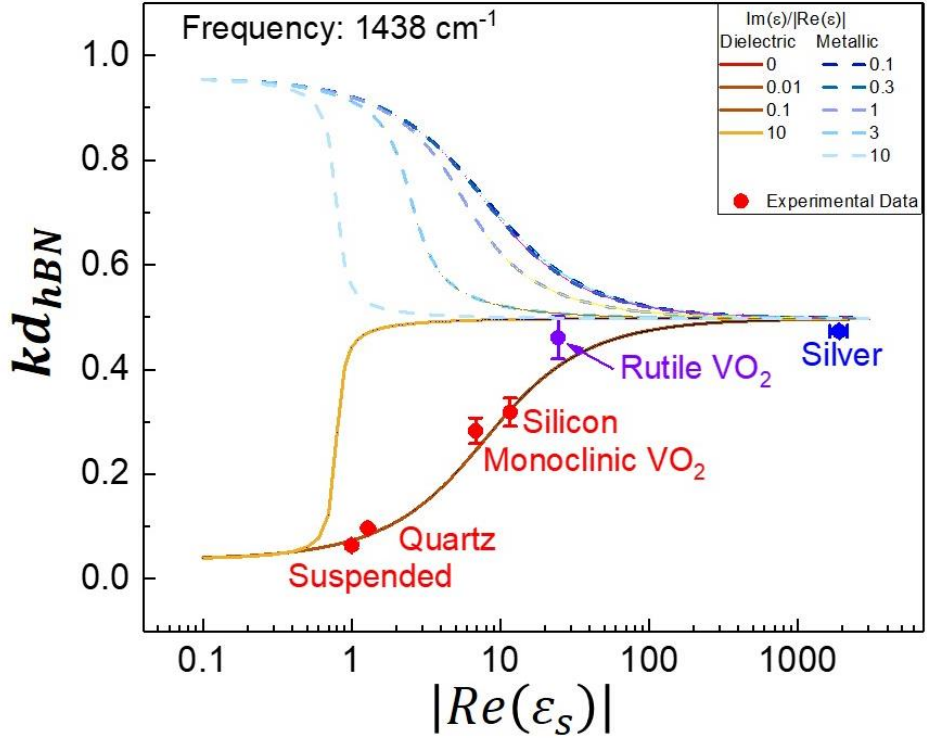

**Figure 3.** The thickness normalized wavevector ($kd_{hBN}$) dependence upon the real permittivity of the substrate dielectric function at 1438 cm$^{-1}$. This dependence is plotted for dielectric (red shaded) and metallic (blue shaded lines) substrates with the shade indicative of the loss tangent (see legend). The red, blue and purple circles are derived from the experimental data presented in this work.

The dispersion relations plotted in Figure 2 and Figure S2 show that the wavelength of the propagating HPhPs can be altered by the choice of substrate and incident frequency, as explored in prior work.[30-32] However, in this study the range of substrate dielectric properties is broader, enabling a systematic examination of the variation of the HPhP wavelength with substrate dielectric function, referenced to the intrinsic properties of suspended hBN. To compare our experimental results and theory, we plot the polariton wavevector as a function of the absolute value of the real part of the dielectric function of the substrate [$|Re(\varepsilon_s)|$] at a single frequency (similar plots at other incident frequencies are provided in the Supporting Information, Figure S6) to draw general conclusions about the dependence of HPhP modes on substrate dielectric function.

To make consistent comparisons between theory and experimental results on hBN flakes of varying thicknesses on different substrates, we make two simplifications. First, we consider the dimensionless HPhP wavevector *kd*, calculated by multiplying the HPhP momentum by the flake thickness, to normalize the dispersion relationship, as in Eq. 1. The thickness dependence of hyperbolic polariton dispersion[30] is a consequence of the Fabry-Perot behavior of hyperbolic modes in thin slabs [11]. Second, we make assumptions about the relative magnitudes of the real and imaginary parts of the substrate dielectric function to account for absorption. Lossless IR dielectrics such as silicon have $Im(\varepsilon_s) = 0$; however for metals and polar materials, the real and imaginary parts of the dielectric function are coupled by the Kramers-Kronig relations. For many metals, the imaginary part of the dielectric function satisfies $0.1 \leq Im(\varepsilon_s)/|Re(\varepsilon_s)| \leq 0.3$,[35] whereas for polar dielectrics $0.01 \leq Im(\varepsilon_s)/|Re(\varepsilon_s)| \leq 0.1$.[21] In certain classes of poor metals, as well as non-crystalline materials, $Im(\varepsilon_s) \gg |Re(\varepsilon_s)|$,[36,37] with $Im(\varepsilon_s)/|Re(\varepsilon_s)|\sim10$ being typical. Thus $Im(\varepsilon_s)/|Re(\varepsilon_s)|$ (the loss tangent) is a good measure of substrate loss properties.

The analytical (Eq. 1 with *l*=0) and experimental results for the normalized wavevector as a function of $|Re(\varepsilon_s)|$ are plotted in Figure 3. A range of analytical curves representing different $Im(\varepsilon_s)/|Re(\varepsilon_s)|$ ratios, for both positive (dielectric) and negative (metallic) $Re(\varepsilon_s)$ at 1438 cm$^{-1}$ are provided. Experimental data are represented as solid points, with error bars derived from the range of dielectric functions for each substrate for the x-axis[35, 38-42] and from measurement uncertainty in the hBN flake thickness and wavevector for the y-axis. (The error estimates are discussed in SI). We compare theory and experiment by considering dielectrics [$Re(\varepsilon_s) >$

1, $Im(\varepsilon_s) \ll Re(\varepsilon_s)$], metals [$Re(\varepsilon_s) < 0, Im(\varepsilon_s) < Re(\varepsilon_s)$] and highly absorbing materials [$Im(\varepsilon_s) > Re(\varepsilon_s)$] separately.

First, we consider dielectric substrates. Analytical results (solid lines) indicate that the HPhP wavevector increases monotonically as a function of substrate permittivity $Re(\varepsilon_s)$, with only minimal influence from substrate-induced absorption loss. Red points indicate experimental data for HPhP modes within hBN on dielectric substrates in this experiment (suspended, quartz, silicon and monoclinic VO$_2$), which match the analytical model. This demonstrates that a high-index substrate has a larger polariton wavevector (smaller wavelength) that is nominally insensitive to substrate loss within the range of typical loss tangents: $Im(\varepsilon_s)/|Re(\varepsilon_s)| < 0.1$. We also emphasize that as $Re(\varepsilon_s)$ tends towards infinity, the value of *kd* tends to a frequency-dependent constant ($\psi$), which can be interpreted using Eq. 1. The only term that includes the substrate dielectric function is $tan^{-1}\left(\frac{\varepsilon_s}{\varepsilon_t \psi}\right)$, which is the phase accumulated upon reflection of the hyperbolic wave from the substrate. As the dielectric function becomes larger, the phase of the reflected polariton tends towards $\pi/2$. We attribute the increase in phase shift to the reduced penetration depth in substrate, with a high real part of the dielectric function, which appears analogous to the Goos-Hänschen shift.

The propagation characteristics of the HPhP modes on metallic substrates are less intuitive. Analytical predictions for metallic substrates are shown as dashed lines in Figure 3. For negative permittivity substrates, the $tan^{-1}$ term in Eq. 1 becomes negative, resulting in a negative shift in the reflected phase. To provide a direct comparison between dielectric and metallic substrates, we set the minimum *l*=1. As the substrate becomes more metallic (i.e.. $Re(\varepsilon_s)$ becomes more negative) the magnitude of the HPhP wavevector is reduced. In the limit of large dielectric constant, the wavevectors of HPhPs in hBN over both metallic and dielectric substrates converge. For metals, this occurs when the plasma frequency is significantly larger than that of the HPhP mode, and thus the complex dielectric constant of the substrate comprises a large negative real permittivity and correspondingly large imaginary index of refraction. From the analytical calculations shown in Figure 3, it would appear at first glance that the dispersion is more sensitive to losses for metallic substrates; however, this is due simply to the higher loss tangents associated with metals. This is consistent with our calculations for dielectrics, as no light can enter any material in which $Re(\varepsilon_s) \ll 0$, and hence experiences a π/2 phase shift upon reflection. In this study, we compared the analytical model to a single noble metal (silver), with the position matching with our calculations.

Finally, we consider a highly absorbing material, represented by rutile (metallic) VO$_2$. The dielectric function of rutile VO$_2$ has not been measured extensively in this frequency range for single crystals.[38, 43] Reference 44 indicates, however, that VO$_2$ is properly classified as a 'bad metal' as it fails to satisfy the Wiedemann-Franz law; therefore, we take it to be an overdamped material with $Im(\varepsilon_s) > |Re(\varepsilon_s)|$. If we assume a large loss tangent, $Im(\varepsilon_s)/|Re(\varepsilon_s)| = 10$, we find that regardless of whether rutile VO$_2$ is considered a bad metal or a lossy dielectric, there is minimal influence upon the normalized wavevector. To highlight the ambiguity in this case, we plot the wavevector of HPhPs supported in hBN on rutile VO$_2$ as a purple circle.

The good agreement between experiment and model clearly indicates the broad applicability of this approach for HPhPs, demonstrating that this approach is indeed capable of reproducing the influence of the substrate, regardless of loss or magnitude or sign of the permittivity in a generalized fashion for all hyperbolic media. The analytical model of Eq. 1 is only appropriate in the large $k$ limit; numerical methods more accurately describe dispersion at small $k$. However, the general trends for both the analytical and numerical models are nearly identical, with the numerical results being required for conditions where $k$ and $2\pi n_s/\lambda_0$ are of the same order, as shown in the Figure S3 and S4 of the Supporting Information. Thus conclusions drawn from the analytical model are representative of the mode behavior in hBN even for thicker flakes.

To investigate the effect of the substrate refractive index on higher-order modes, we again plot the analytical dispersion relation in Figure 4, including now the second- and third-order modes at 1510 cm$^{-1}$ and a loss tangent of 0.01 (0.3) for dielectric (metallic) substrates. These higher-order branches correspond to shorter-wavelength polariton modes at the same incident frequency[6, 12] (Figure 2). As with the principal HPhP mode, these high-order polaritons are also affected by the substrate dielectric function; however, due to the reduced range of the evanescent field associated with the shorter polariton wavelengths, one would expect this effect to be significantly reduced. However, the change in wavevector $\Delta k = k(\varepsilon_{s1}) - k(\varepsilon_{s2})$, from a low $\varepsilon_s$ substrate with wavevector $k(\varepsilon_{s1})$ to a high $\varepsilon_s$ substrate with wavevector $k(\varepsilon_{s2})$ is actually the same for all hyperbolic modes. Naively, this suggests that the substrate will have the identical influence on all higher-order modes in the hyperbolic material, with no significant change in the properties of different modes other than the degree of volume confinement.

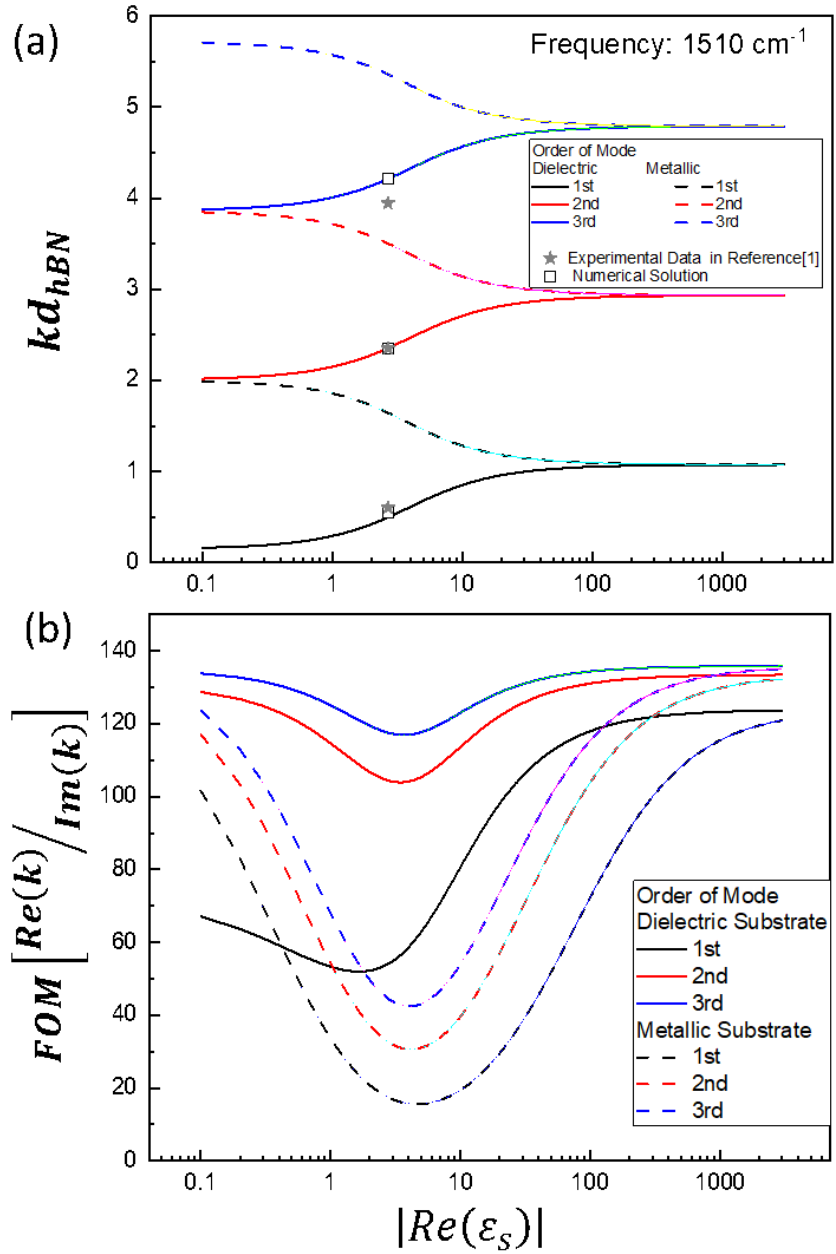

**Figure 4.** Influence of substrate dielectric function upon high-order HPhP modes of hBN at 1510 cm$^{-1}$, for dielectric (metallic) substrates, with the loss tangent fixed at 0.01 (0.3). a) $k \cdot d$ was plotted versus $|Re(\varepsilon_s)|$, and gray stars are experimental data in Ref. 12, while open black boxes are numerical solutions for 120 nm thick hBN. b) FOM of the first three orders of HPhPs in this same thickness flake of hBN as a function of the $|Re(\varepsilon_s)|$.

However, the *fractional* change in wavevector $\frac{k(\varepsilon_S)}{k(\varepsilon_{S\prime})} = 1 + \frac{\Delta k}{k(\varepsilon_S)}$ is often a more useful metric in comparing HPhP propagation, where $\varepsilon_{S\prime}$ refers to a second substrate to which the first is compared. This is because phenomena such as refraction are entirely dependent on the fractional change in wavector. For higher-order modes, with larger initial values of $k$, this fractional change becomes smaller, approaching unity—i.e. the higher-order modes are less sensitive to substrate dielectric function. For example, between air and silicon (at 1510 cm$^{-1}$) the wavelength changes by factors of 7.4 and 3.1 for the first- and second-order modes, respectively. This has significant implications for the behavior of HPhP propagation across substrate boundaries; for example, lower-order modes are more strongly refracted at such boundaries, an effect which could be exploited to separate the different modes as demonstrated below (Figure 5). While the experimental methods deployed here precluded observation of higher-order modes due to the thinness of the hBN flakes - chosen to ensure the validity of the analytical model – we have added data from Ref. 12 as well as numerical solutions from the model in Ref. 11 to validate these conclusions (Figure 4a). The experimental data agree well with predicted mode positions, with slight deviations for the third-order mode attributed to the challenges associated with launching high wavevector modes and the concomitant error in extracting the polariton wavelength given the correspondingly short propagation lengths.

To compare propagation properties of HPhP modes on different substrates we use a figure of merit (FOM)[12] related to the real and imaginary parts of the wavevector:

$$FOM = \frac{Re(k)}{Im(k)} \qquad (2)$$

This FOM is better than the propagation length as it to accounts for the stronger polaritonic waves confinement typical of correspondingly shorter propagation lengths. Furthermore, while long propagation lengths can be realized for weakly confined polaritons, it is only within the limit of strong modal confinement that the intrinsic benefits of sub-diffractional wavelengths associated with polaritons can be exploited. Specifically, $Im(k)$ determines the propagation length $L_p = \frac{1}{2 \cdot Im(k)}$,[30] and $Re(k)$ defines the polariton wavelength $\lambda = \frac{2\pi}{Re(k)}$, and thus the FOM can be expressed as:

$$FOM = \frac{2\pi}{\lambda} \times 2L_p = 4\pi Q \qquad (3)$$

where $\lambda$ is polariton wavelength, $L_p$ is propagation length, and Q is number of cycles of the polariton wave before the amplitude decreases to 1/$e$ of its initial value. The relationship between this FOM and the substrate dielectric function is plotted in Figure 4b, using the loss tangents from Figure 4a.

From the general trends, the FOM clearly decreases rapidly with increasing substrate dielectric constant, reaching a minimum at approximately |ε$_s$| = 3, and then increasing again. This effect is especially pronounced for the principal mode on dielectric substrates, as high-index substrates yield stronger confinement and longer propagation lengths. This contrasts with, and goes well beyond, prior results,[30] which showed only that FOM increased for suspended hBN compared to hBN on silicon. For metallic substrates, the effect is similar; however, the FOM is approximately symmetric about the minimum with respect to the dielectric response. Interestingly, this implies that when the substrate permittivity is close to the epsilon-near-zero (ENZ) condition rather than a metallic or dielectric value, the FOM for lossy substrates might actually be higher than that for a low-loss dielectric substrate.

For both dielectric and metallic substrates, the FOM of the higher-order modes is always improved with reference to the principal mode, though the trends with substrate permittivity are generally similar. Overall, the results of Figure 4 indicate that the choice of substrate is a complicated issue for hyperbolic polaritons. While low-index substrates produce long-wavelength propagating modes, very high index substrates actually produce the highest FOMs for those modes. In the Supporting Information we discuss the effect of the substrate loss tangent on the first and higher-order modes, (see Figure S5).

These results have significant implications for device designs based on HPhPs, in particular for both polariton refraction in planar metasurface-based optics and hBN-based sensors. Previously we showed that HPhPs refract when propagating across a boundary between metallic and dielectric regions of a phase-change material.[2] The fact that each successively higher-order mode exhibits a smaller change in wavelength due to the changing dielectric environment enables spatial sorting of mode orders. To illustrate the potential for using local dielectric function to control HPhP propagation, we consider a hypothetical device capable of spatially separating different mode orders by refraction at dielectric boundaries (Figures 5a and b). The simulations are configured with a 120 nm thick hBN flake that is partly suspended and partly supported on a Si substrate. The HPhPs are launched from an array of dipole emitters above the suspended hBN, with the HPhPs propagating toward the air-Si interface at a 45° angle of incidence. To suppress HPhP reflections from the simulation boundary, we surround the region of interest by a border of highly lossy (20x increased damping) hBN. The *z*-component of the electric field at a position 10 nm above the hBN surface is provided to demonstrate the varied refraction of principal and higher-order modes leading to spatial separation (Figure 5b). The HPhPs launched from the emitter array are refracted at the air-Si interface due to the wavevector mismatch across the boundary, but since the principal mode is significantly more sensitive to the surrounding dielectric environment, it is more strongly refracted (transmitted angle 13°, orange arrow) and is thus spatially separated from the second-order mode (transmitted angle 34°, green arrow). Therefore, using such approaches, high wavevector, higher-order HPhPs could be spatially separated from the longer wavelength principal modes in a planar film.

To discuss the implications for index-based sensing,[45] where the change in frequency of polaritonic resonance indicates a change in the local dielectric environment, we consider the analytical model for both surface phonon polaritons (SPhPs) and HPhPs. At mid-IR frequencies, strongly confined HPhPs and environmental sensitivity in hBN-enabled surface-enhanced infrared absorption (SEIRA)[46-48] spectroscopy and suggested a basis for planar metaoptics structures[2] that have since reduced to practice.[17] These new results also have implications for deploying hBN in thin-film sensing applications. Thin film sensing operates in two main regimes: index-based sensing (which relies on the change in local dielectric environment to change the mode frequency), or surface enhanced infrared absorption (which exploits local field enhancements). Our results are most obviously relevant in the field of index-based sensing, as they consider the properties of a substrate without significant spectral dispersion. As shown in Figure 3, the presence of a nearby analyte significantly affects the properties of suspended hBN – resulting in a measurable shift in resonance frequencies. Thus, it is critical that hBN be suspended or on a low-permittivity substrate to function as an effective sensor (see Supporting Information Figure S7).

To demonstrate how such a device could be realized, a device consisting of 120 nm thick hBN atop a 100 nm thick gold structure surrounding a resonant cavity filled with air or Si is simulated in Figure 5b. Since HPhPs in hBN are sensitive to the surrounding dielectric environment, the resonant mode of this structure will be sensitive to the presence of an analyte on top of the hBN layer. Figures 5d and e show simulated reflectance spectra for an array of such resonators under far-field radiation incident at 45°, *p*-polarized light. The structure is covered with a hypothetical analyte layer of $Re(\varepsilon)=2.25$ of variable thickness. For both air- and Si-filled resonators (Figure 5d and e, respectively), sharp dips appear corresponding to resonant absorption in the cavity defined by the gold. These resonant frequencies are sensitive to as little as 1 nm of analyte placed on the hBN. However, using suspended hBN significantly increases the magnitude of the observed shifts to well over 10 cm$^{-1}$.

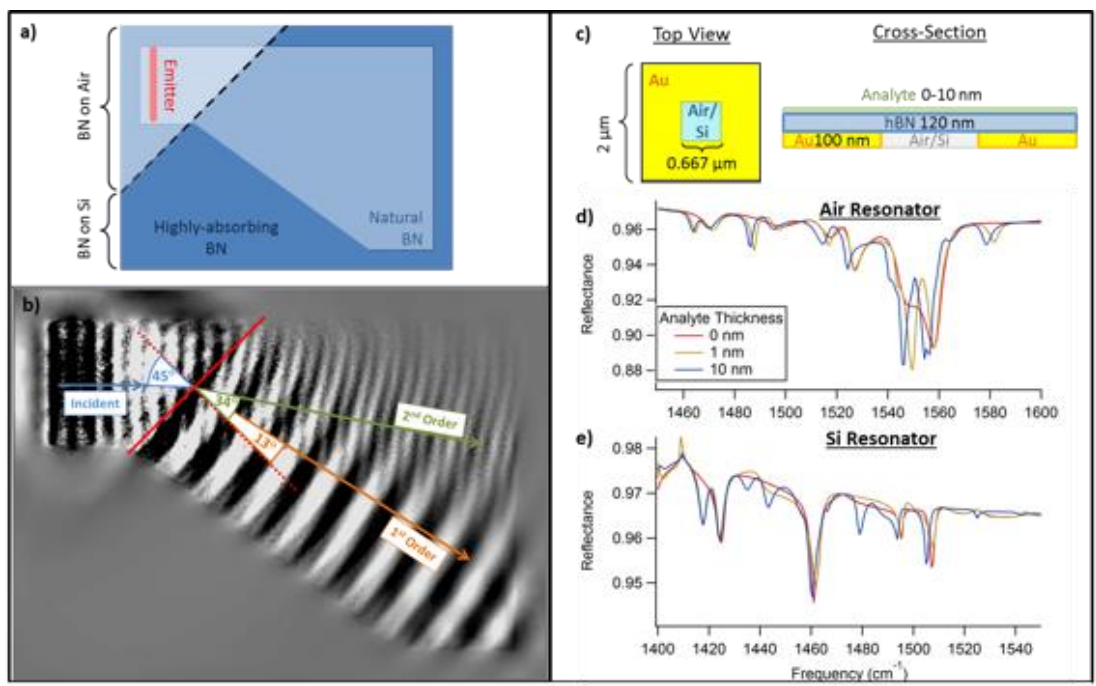

**Figure 5**: Simulations of spatial separation of the HPhP modes and applications of HPhPs through dielectric enviornment control. a) A schematic diagram of the proposed HPhP spatial separator device, with the corresponding b) simulated electric field ($E_z$) for HPhP modes separated due to refraction resulting from HPhP transmission over a Si-air interface aligned at 45º with respect to the propagation direction. To eliminate complexities in the image due to back reflections, a notional highly-absorbing hBN border with a 20-fold increase in the damping constant was added. The corresponding refraction of the principal (orange) and 2$^{nd}$ order (green) HPhPs are provided. c) Schematic diagram of a proposed hBN HPhP-based SEIRA or index-sensing resonant device induced due to large index contrast in spatial regions underneath the hBN flake as designated. The corresponding reflection spectra for this resonant structure fabricated from a hBN flake over an empty d) or Si-filled e) resonant chamber (e.g. hole). The calculated resonance spectra are highly sensitive to the presence and thickness of an analyte layer on top of the hBN surface; suspended hBN films have the highest degree of sensitivity. The spectra provided for both cases in d) and e) as a function of analyte layer thickness are labeled.

Whilst this simulation suggests that hBN can act as a surface sensor, it is instructive to benchmark against other index-based sensing schemes. In the Supporting Information (Figures S8 and S9) we compare a hBN film against a notional isotropic SPhP material operating with the same TO, LO and damping frequencies. Surprisingly, we find that for thick films of the analyte material ($>\lambda_{HPhP}$), the SPhP mode is much more effective as an index sensor. However, taking into account the strong confinement of HPhPs in hBN, we find that hyperbolic modes are indeed much more effective for the analysis of thin films, especially close to the LO phonon (where HPhP modes are extremely confined). This is reflected in the results of Figure 5d, where the peak shifts are largest close to the LO phonon energy.

We have investigated the interaction of HPhP with substrates encompassing a wide range of complex refractive indices. While substrates with small real parts of the dielectric function support long-wavelength propagating modes, large permittivity substrates result in polaritons exhibiting the highest propagating FOMs. Furthermore, our results demonstrate that longer wavelength principal hyperbolic modes can be used as extremely sensitive subwavelength sensors. While all modes show an equal change in wavevector ($\Delta k$), the corresponding fractional change in wavevector $k(\varepsilon_{s1})/k(\varepsilon_{s2})$ for higher-order modes implies that they are less influenced by the local dielectric environment than the principal. Most significantly, the substrate permittivity can induce spatial mode separation of these higher-order modes. This effect could be used as a tool to optimize planar refractive optics and reconfigurable metasurfaces. Our results therefore provide a deeper understanding of HPhP interactions with the surrounding environment, a necessary step for implementing practical applications in on-chip molecular sensing and nanophotonics.

**Methods** *Sample Preparation.* For the purposes of these experiments, we employed isotopically pure hBN flakes (>99% h[10]BN) to minimize the intrinsic polariton losses.[12, 49] These were grown as described in reference,[50] then subsequently exfoliated and transferred onto the appropriate substrate. Single crystals of $VO_2$ were grown on quartz from vanadium pentoxide powder ($V_2O_5$) by physical vapor transport.[51] Due to the presence of multiple $VO_2$ crystals on the substrate we were able to suspend the hBN between adjacent $VO_2$ crystals. This resulted in flakes which were supported by $VO_2$ and quartz or alternatively suspended between $VO_2$ crystals (as shown in Figure 1). Silver films were deposited on Si substrates, while $SiO_2$ and Si substrates were obtained commercially.

*s-SNOM.* In s-SNOM a platinum-coated probe tip is used both to map the topography and to probe the optical near fields. Monochromatic infrared near field imaging at selected laser excitation frequencies is performed via a combination of phase interferometric detection and demodulation of the detector signal at the second harmonic ($2\Omega$) of the tip oscillation frequency.[52] The nano-FTIR data were acquired using a combination of s-SNOM and a broad-band infrared light source (neaspec.com).

AUTHOR INFORMATION

**Corresponding Author**

Yohannes Abate. E-mail: yabate@physast.uga.edu

**Author Contributions**

The manuscript was written through contributions of all authors. All authors have given approval to the final version of the manuscript. Y.A., R.F.H., and J.D.C. conceived and guided the experiments. S.T.W. grew the $VO_2$ crystals and identified the phase domains. S.L. and J.H.E. grew the hBN crystals. T.G.F. and S.T.W. fabricated the hBN-$VO_2$ heterostructure. A.F. and N.A. performed s-SNOM and Nano-FTIR experiments and T.G.F., M.H., S.T.W., A.F. and N.A.

analyzed the data. M.H., S.T.W. and T.G.F. provided the analytical and finite element simulations included in Figures 3, 4 and 5. All authors contributed to writing the manuscript.

**Funding Sources**

Y.A. and N.A. gratefully acknowledge support provided by the Air Force Office of Scientific Research (AFOSR) grant number FA9559-16-1- 0172. The work of A.F. is supported by the National Science Foundation grant 1553251. Support for the $^{10}$B-enriched hBN crystal growth was provided by the National Science Foundation, grant number CMMI 1538127.

**Notes**

The authors declare no competing financial interest.

ACKNOWLEDGMENT

We thank Professor Misha Fogler for providing a script to calculate the dispersion of HPhPs. T.G.F. and S.T.W. thank the staff of the Vanderbilt Institute for Nanoscience (VINSE) for technical support during fabrication and Kiril Bolotin for preliminary design of the 2D transfer tool used.

ABBREVIATIONS

HPhP, Hyperbolic phonon polariton; s-SNOM, scattering type scanning near field microscopy; FTIR, Fourier transform infrared; IR, Infrared; PCM, phase change material; $VO_2$, Vanadium dioxide; hBN, hexagonal boron nitride; FoM, Figure of Merit.

# Refractive Index-Based Control of Hyperbolic Phonon-Polariton Propagation

*Alireza Fali[†], Samuel T. White[‡], Thomas G. Folland[∥], Mingze. He[∥], Neda A. Aghamiri[†], Song Liu[⊥], James H. Edgar[⊥], Joshua D. Caldwell[∥, δ], Richard F. Haglund[‡, δ], Yohannes Abate[*,†]*

[†] Department of Physics and Astronomy, University of Georgia, Athens, GA 30602, United States

[‡] Department of Physics and Astronomy, Vanderbilt University, Nashville, TN 37235, United States

[∥] Department of Mechanical Engineering, Vanderbilt University, Nashville, TN 37212, United States

[⊥] Tim Taylor Department of Chemical Engineering, Kansas State University, Manhattan, KS 66506 USA

[δ] Interdisciplinary Materials Science Program, Vanderbilt University, Nashville, TN 37212

# SUPPLEMENTAL INFORMATION

## S1: Additional example of HPhP wavelength modulation across multiple substrates

Another exfoliated hBN flake draped over a $VO_2$ single crystal, here on top of a Si substrate, is presented in Figure S1. Again, we observe HPhPs of different wavelengths in hBN in three different environments: on Si substrate, suspended above the substrate, and on $VO_2$. As discussed in the main text, the dielectric properties of the surrounding environment modulate the HPhP propagation in hBN.

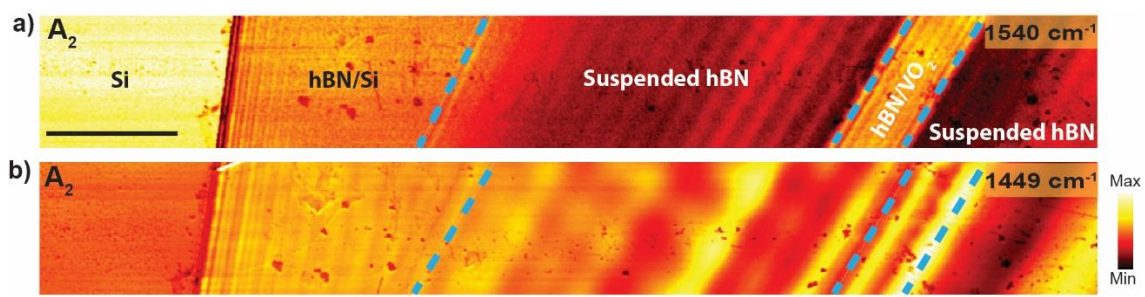

Figure S1: Substrate-dependent modulation of the polariton wavelength in an hBN flake draped over a Si substrate and $VO_2$ crystal. (a) and (b) HPhP amplitude near field images taken at 1540 cm$^{-1}$ and 1449 cm$^{-1}$, respectively. Scale bar is 5 µm.

## S2: SNOM images and polariton dispersion on silver and silicon substrates

Figure 2 of the main text shows calculated and experimental dispersion relations for HPhPs in hBN suspended in air, on quartz, or on metallic or insulating $VO_2$. The corresponding data for the two other substrates discussed in the paper (silicon and silver) are shown in Figure S2. Solid curves in the dispersion plots [Figure S2(a) and (d)] are calculated from the analytical formula Eq. 1; solid blue circles are experimental data points derived from s-SNOM images such as those presented in Figure. S2 (b-c) and (e-f).

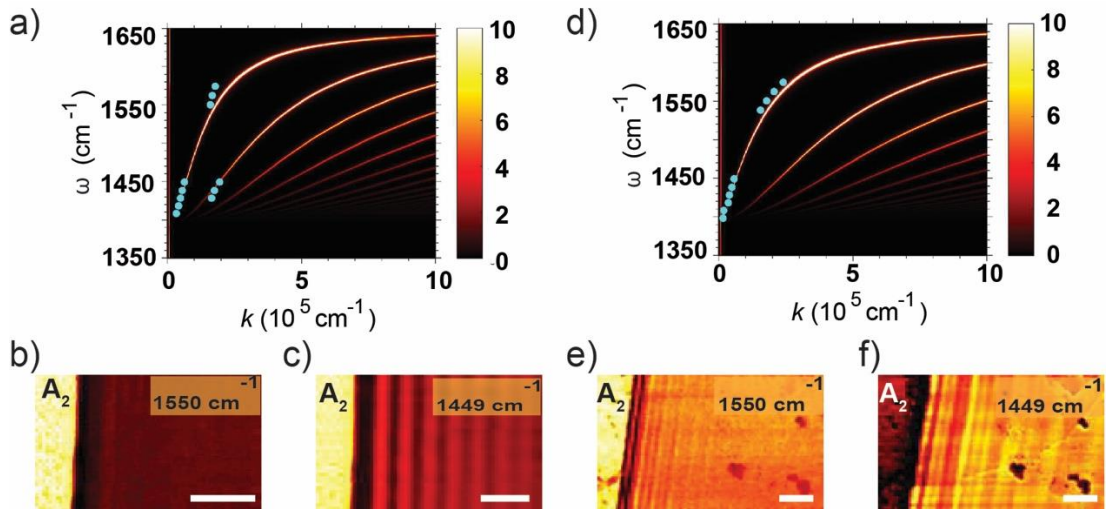

Figure S2: Dispersion relation of HPhP of hBN: (a) a 85 nm thick hBN flake on silver (b) and (c) HPhP amplitude near field images taken at 1550 cm$^{-1}$ and 1449 cm$^{-1}$. (d) a 65 nm thick hBN flake on silicon (e) and (f) HPhP amplitude near field images taken at 1550 cm$^{-1}$ and 1449 cm$^{-1}$. Experimental data points (extracted from monochromatic) are shown superimposed on analytical calculation. Scale bar is 1 µm.

## S3: Systematic error in the analytical calculations

As mentioned in the main text discussion regarding Figure 3, the analytical solution is based on the approximation that $k$ is much larger in hBN than in the surrounding environment (i.e. substrate and air), so that $k$ can be neglected in those environments. Numerical simulations, on the other hand, do not rely on such assumptions. When the hBN is sufficiently thin (e.g. 20 nm), the field is highly confined, $k$ is large, and the analytical and numerical solutions are identical (Figure. S3, open red boxes and red line). For thicker hBN (e.g. 100 nm), where $k$ is smaller at a given frequency, the analytical solution is less accurate: it yields a $k$ lower than the experimental value[1] (grey star), whereas the numerical simulation agrees well (Figure S3, open grey boxes).

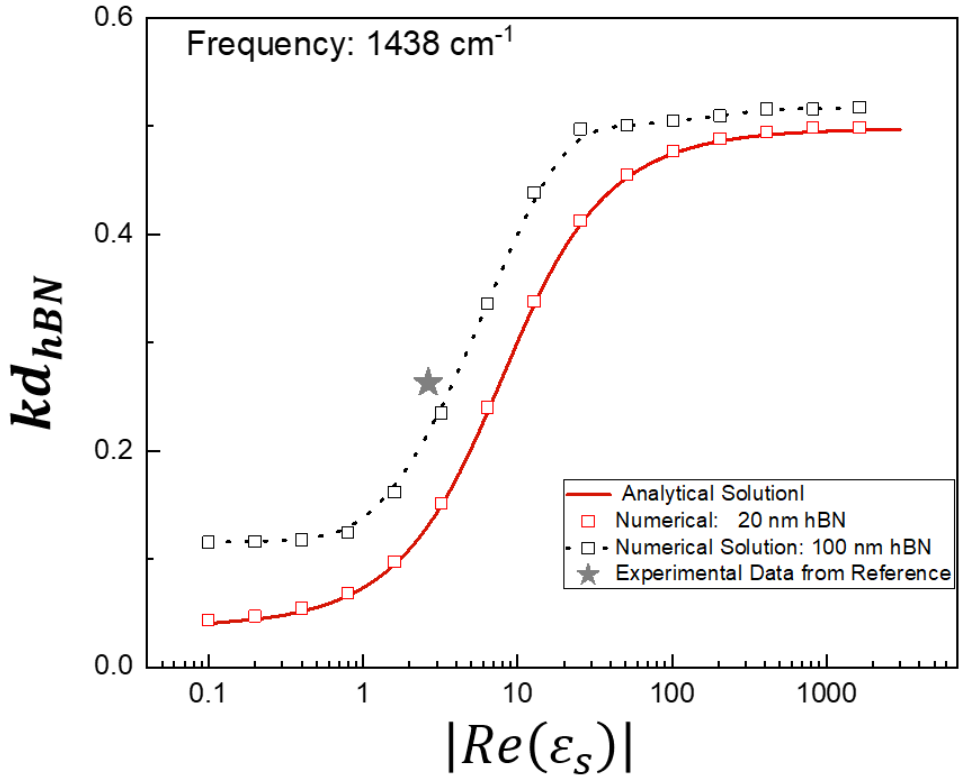

Figure S3: The thickness-normalized wavevector $kd$ versus substrate dielectric function, at 1438 cm$^{-1}$, for dielectric substrates $[Im(\varepsilon_s)/Re(\varepsilon_s) = 0.01]$. Red line is the analytical solution. Red(grey) open boxes are numerical solutions for 20 nm (100 nm) hBN. Grey star is an experimental datum from reference [1].

This error is more pronounced at low frequencies, where $k$ is small, but disappears as frequency increases and $k$ becomes arbitrarily large, even for thick hBN. Figure S4 illustrates this behavior by plotting the thickness-normalized dispersion curves for suspended and Si-supported hBN. For thin (20 nm) hBN, the analytical and numerical solutions agree at all frequencies; for thick (120 nm) hBN, they agree at high frequencies, but diverge as the frequency decreases, the analytical solution consistently yielding a too-small $k$ value. Thus, the analytical approximation holds well as long as the hBN is thin, or the frequency is high.

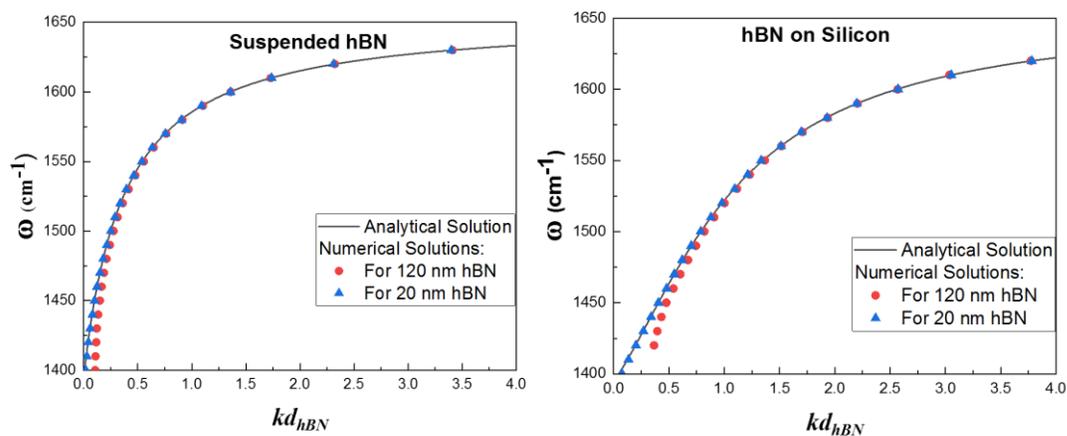

Figure S4. Comparison of numerical and analytical solutions as a function of frequency, for different hBN thicknesses, suspended (left) and on Si (right). The thickness-normalized wavevector $kd$ is plotted to suppress the influence of hBN thickness. Blue (red) triangles (circles) are numerical solutions at 20nm (120nm). The analytical solutions (solid lines) diverge from the numerical solutions in thick flakes at low frequencies.

## S4: The dependence of FOM on substrate loss

Figure S5 illustrates the dependence of FOM on the substrate loss tangent, considering the same values of the loss tangent as in the main text, Figure 2. For loss-free dielectric substrates [$Im(\varepsilon_s) = 0, Re(\varepsilon_s) > 0$], the FOM increases monotonically with $Re(\varepsilon_s)$. For lossy dielectrics, on the other hand, the FOM decreases to a minimum and then increases, asymptotically approaching a limiting value at high $Re(\varepsilon_s)$. For metallic substrates [$Re(\varepsilon_s) < 0$], the trend is similar to that for lossy dielectrics, except that in metallic substrates the low-$Re(\varepsilon_s)$ limiting value of FOM is higher than in dielectrics. As a result, for epsilon-near-zero (ENZ) materials, the FOM of hBN is higher on metallic substrates (despite their greater loss), due to the larger $k$ (see main text Figure 4(a)).

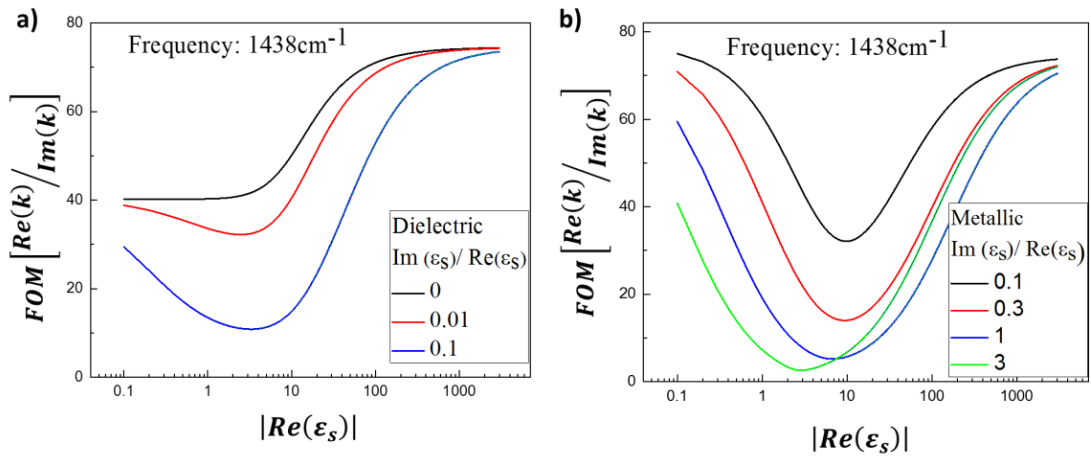

Figure S5: FOM of first-order HPhPs in hBN on (a) dielectric ($Re(\varepsilon_s) > 0$) and (b) metallic ($Re(\varepsilon_s) < 0$) substrates with different loss tangents at 1438 cm$^{-1}$, based on the analytical solution.

Generally, the FOM decreases as the loss tangent increases, though for high levels of loss, the FOM minimum shifts, so that for a given value of $Re(\varepsilon_s)$, it is possible for FOM to increase with loss tangent. In summary, the relationship between $\varepsilon_s$ and the HPhP FOM can be complex and non-intuitive, and these computations will help identify the ideal substrate for a given purpose.

## S5: Standard deviation analysis of dielectric function

|  | Re($\varepsilon_s$) | STD of Re($\varepsilon_s$) | Im($\varepsilon_s$) | STD of Im($\varepsilon_s$) | Loss Tangent | STD of Loss Tangent |
|---|---|---|---|---|---|---|
| Silver[2-7] | $-2.58 \times 10^3$ | 58 | 590 | 104 | 0.228 | 0 |
| Quartz[8, 9] | 1.23 | 0.06 | 0.006 | 0.004 | 0.005 | $2.0 \times 10^{-5}$ |
| Rutile VO$_2$ [10] | 13 | 15 | 91 | 17 | 7.1 | 2.5 |
| Monoclinic VO$_2$ [10] | 6.67 | 0.43 | 0.33 | 0.16 | 0.05 | $6.0 \times 10^{-4}$ |

Table S1: Values and standard deviations for dielectric functions of different substrates at 1438 cm$^{-1}$, used in Figure 3 of the main text. Dielectric functions are averaged from values drawn from literature, with standard deviations representing differences across reports.

In Figure 3 of the main text, x-axis error bars represent the standard deviation (STD) in $\varepsilon_s$ of the substrates as reported in the literature, the error arising from inconsistencies across these reports. The values of $\varepsilon_s$ for silver[2-7] and quartz[8, 9] in this frequency range have been reported several times with little deviation, as shown in Table S1. The silicon substrate we used is diced from wafer-size silicon grown at foundry level, and should have negligible error.

However, for VO$_2$, especially for rutile VO$_2$, the dielectric functions from literature are inconsistent. Reference [10] reported refractive index and extinction coefficient of four different VO$_2$ films, grown by different methods and on Si and sapphire substrates. As VO$_2$ grown on sapphire is very different from that grown on Si, we use only the data sets for VO$_2$ grown on Si wafer. For monoclinic VO$_2$, $\varepsilon_s$ is 6.67 with an STD of 0.43, or merely 6%. In rutile VO$_2$, on the other hand, the variation in $\varepsilon_s$ is so large we cannot be sure whether it is metallic [$Re(\varepsilon_s) < 0$] or dielectric [$Re(\varepsilon_s) > 0$]—at this frequency, rutile VO$_2$ is either an extremely lossy metal or extremely lossy dielectric. However, this would have negligible influence on the HPhP dispersion, as mentioned in the main text.

## S6: Substrate dependence for different frequencies

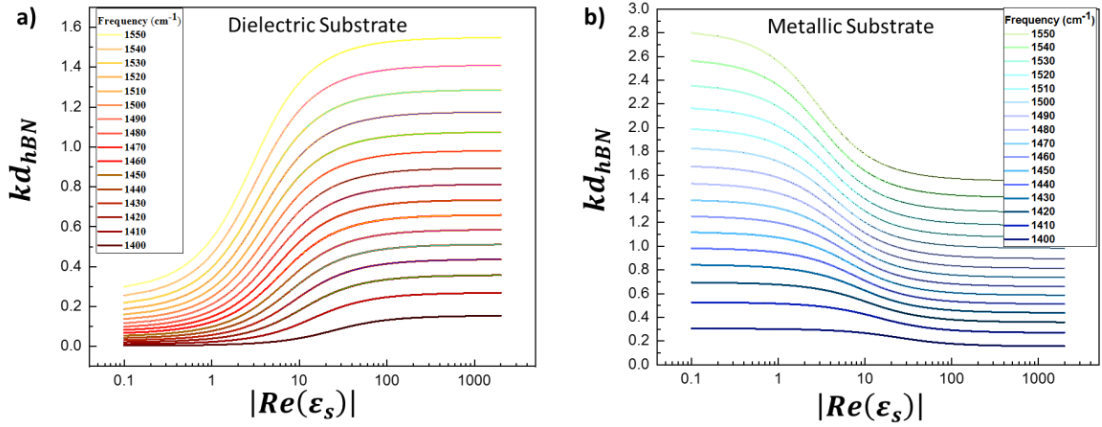

Figure S6: The relationship between $kd$ and $Re(\varepsilon_s)$ at different frequencies for (a) dielectric substrates, and (b) metallic substrates.

As mentioned in the discussion of Figure 3 in the main text, the normalized wavevector $kd$ and its change as a function of $Re(\varepsilon_s)$ is sensitive to the frequency of the polariton mode. To show the influence of polariton mode frequency we plot the kd over $[Re(\varepsilon_s)]$ for a range of frequencies (Figure S6). Generally speaking, at higher frequencies the HPhP is more sensitive to the local dielectric environment, resulting in a larger change in the polariton wavelength. This can be attributed to the change in the HPhP propagation angle within the hBN (main text Ref. 5,15), which results in a stronger interaction with the surrounding dielectric environment.

## S7: Analytical analysis of HPhP vs SPhP sensing on a semi-infinite half space

For index-based surface sensing experiments, we wish to assess how much the frequency changes ($\partial \omega$) when exposed to a small change in the dielectric function of an analyte material on top of the hBN ($\partial \varepsilon_t$). We treat the analyte as semi-infinite, which allows us to use the dispersion relation of Eq (1). The result will be inherently dependent on the substrate dielectric function $\varepsilon_s$, as well as the incident frequency and the initial value of $\varepsilon_t$. To calculate $\partial \omega / \partial \varepsilon_t$, we write $\partial \omega / \partial \varepsilon_t = (\partial \omega / \partial k) \cdot (\partial k / \partial \varepsilon_t)$, where the derivative $(\partial \omega / \partial k)$ represents the group velocity, and $(\partial k / \partial \varepsilon_t)$ represents the change in wavector upon a change in the dielectric environment.

Results from this calculation are presented in Figure S7(a). If the hBN is suspended ($Re(\varepsilon_s) = 1$), the change in frequency could be as high as 32 cm$^{-1}$ per unit change in $\varepsilon_t$. If $\varepsilon_t$ starts as a higher value, this sensitivity is reduced. Crucially, $\partial \omega / \partial \varepsilon_t$ is always much higher for hBN suspended than on substrates like Si, by nearly an order of magnitude. Furthermore, if the analyte surrounds hBN rather than lying on one side only, $\partial \omega / \partial \varepsilon_t$ can be nearly doubled, as shown in Figure S7 (b). These calculations illustrate that hBN on a low-$Re(\varepsilon_s)$ material is preferable for sensing purposes (as noted in the main text). We note that the parameter $\partial \omega / \partial \varepsilon_t$ is invariant on the thickness of the hBN film, as discussed below.

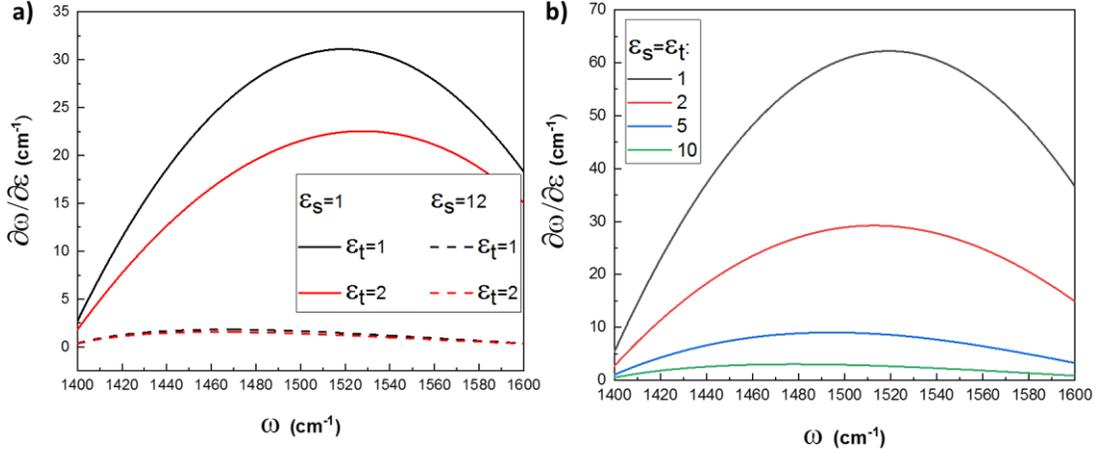

Figure S7: The sensitivity of HPhPs to surrounding dielectric environment for index sensing. The rate of change of $\omega$ with respect to $Re(\varepsilon_t)$ is plotted over incident frequency for (a) suspended or Si-supported hBN covered with analyte or (b) hBN surrounded by analyte.

To compare this index sensing to prior work, particularly to that employing surface phonon polaritons (SPhP), we simulate an artificial 'isotropic hBN', with the same phonon energies along all crystal axes, and perform a similar analysis to that presented above. With this assumption, we compare $\partial\omega/\partial\varepsilon_t$ for both SPhP and HPhP systems used to sense a material with $[Re(\varepsilon_s)]$ close to 1 (Figure S8). Surprisingly, the max $\partial\omega/\partial\varepsilon_t$ of SPhP modes is almost 10 times higher than that of HPhP modes, as shown in figure S8(a). To understand this difference, we plot the two constituent derivatives $\partial\omega/\partial k$ and $\partial k/\partial\varepsilon_t$, for both SPhP [Figure S8(b)] and HPhP [Figure S8(c)] systems. Examining $\partial k/\partial\varepsilon_t$ (red curves), we find that $k$ is much more sensitive to $\varepsilon_t$ in HPhPs than in SPhPs. However, due to the highly dispersive nature of the HPhPs, and consequently low group velocity, of hyperbolic materials, $\partial\omega/\partial k$ (black curves) is much smaller. Overall, $\partial\omega/\partial\varepsilon_t$ is smaller in the HPhP material than in the SPhP material.

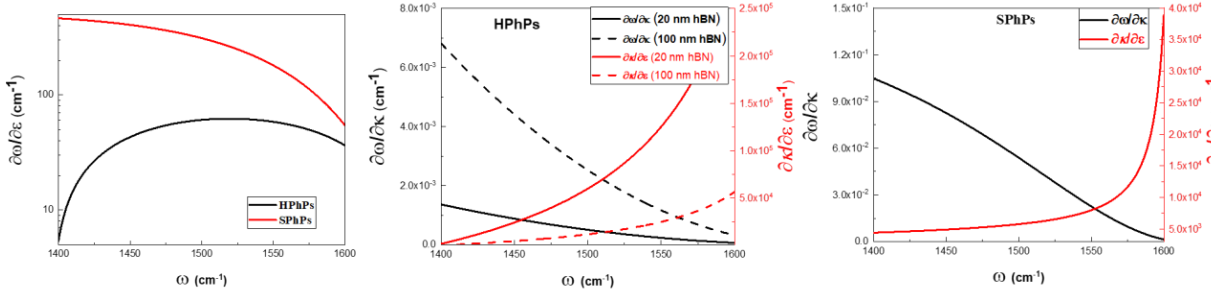

Figure S8: Comparison of sensing benchmark for SPhPs and HPhPs, assuming that the analyte material is semi-infinite with $Re(\varepsilon_s) = 1$. (a) $\partial\omega/\partial\varepsilon_t$ for 20-nm thick hBN on air, and a hypothetical isotropic SPhP-supporting material; (b) $\partial\omega/\partial k$ and $\partial k/\partial\varepsilon_t$ for hBN on air, 20-nm (solid line) or 100-nm (broken line) flakes; (c) $\partial\omega/\partial k$ and $\partial k/\partial\varepsilon_t$ for the SPhP material.

Note, while we chose 20-nm thick hBN for comparison to the SPhP system, the result is valid for other hBN thickness. For example, in 100-nm hBN [dashed lines in Figure S3(b)] $\partial\omega/\partial k$ is 5 times larger (its dispersion is smaller, $v_g$ is higher); but $\partial k/\partial\varepsilon_t$ would be 5 times smaller, since $kd$ is fixed for a given surrounding environment and frequency. Thus, the changes in $\partial\omega/\partial k$ and $\partial k/\partial\varepsilon_t$ due to thickness variation would compensate for each other, leaving $\partial\omega/\partial\varepsilon_t$ unchanged.

However, the above discussion assumes that the analyte is semi-infinite; for analyte films which are only a few atomic layers thick, the relatively low confinement of SPhPs away from the LO-phonon energy could lead to reduced sensitivity. To account for the improved confinement of hyperbolic polaritons, we compute $\partial\omega/\partial\varepsilon_t$ multiplied by the polariton wavevector (Figure S9). Due to the significantly enhanced confinement of HPhPs, they show improved sensitivity close to the LO-phonon energy. This is particularly pronounced for thin flakes of hBN, illustrating that a thin hyperbolic material may offer the best sensitivity for index-based sensing applications.

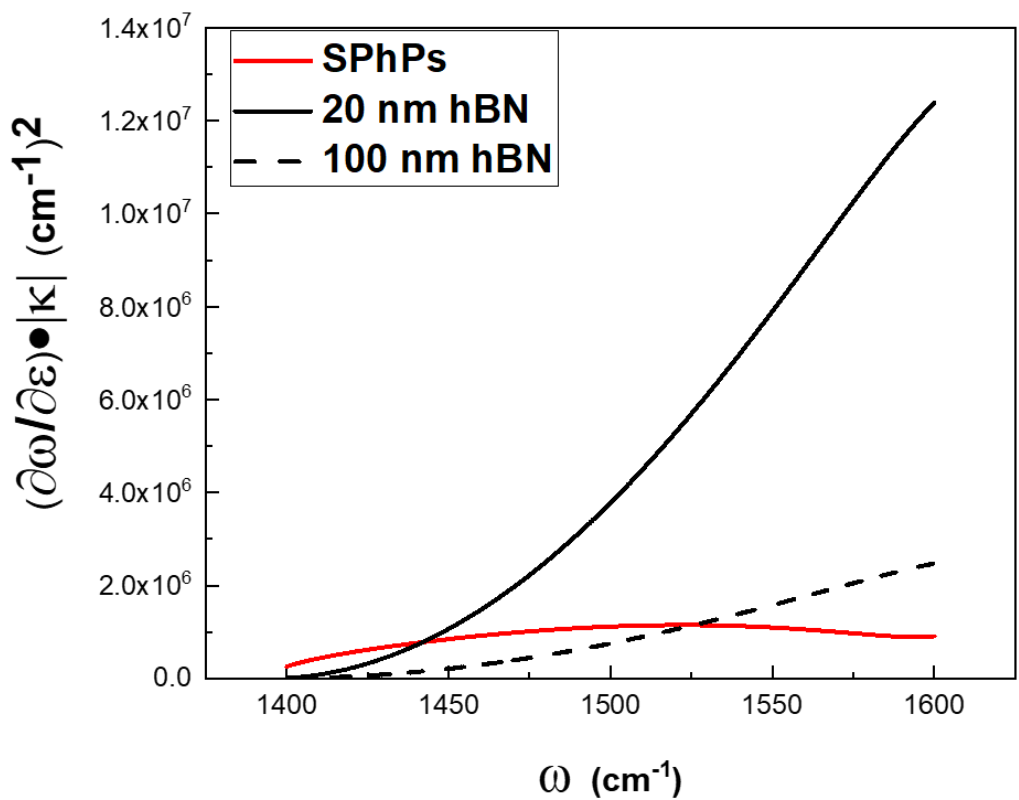

Figure S9. Confinement-normalized sensitivity to dielectric environment. SPhP modes (red line) are more sensitive at low frequencies, but HPhP modes (black lines) become more sensitive as the frequency approaches that of the LO-phonon.